\documentclass{aa}

\usepackage{psfig,graphics}
\usepackage{latexsym}

\def\ros{{\sl ROSAT}}
\def\etal{{et\,al.}}

\def\it{\sl}
\def\degs{\ifmmode ^{\circ}\else$^{\circ}$\fi}
\def\amin{\ifmmode ^{\prime}\else$^{\prime}$\fi}
\def\asec{\ifmmode ^{\prime\prime}\else$^{\prime\prime}$\fi}
\def\farcs{\hbox{$.\!\!^{\prime\prime}$}}  

\newbox\grsign \setbox\grsign=\hbox{$>$}
\newdimen\grdimen \grdimen=\ht\grsign
\newbox\laxbox \newbox\gaxbox
\setbox\gaxbox=\hbox{\raise.5ex\hbox{$>$}\llap
     {\lower.5ex\hbox{$\sim$}}}\ht1=\grdimen\dp1=0pt
\setbox\laxbox=\hbox{\raise.5ex\hbox{$<$}\llap
     {\lower.5ex\hbox{$\sim$}}}\ht2=\grdimen\dp2=0pt
\def\gax{\mathrel{\copy\gaxbox}}

\begin{document}

\voffset=15mm

   \thesaurus{06         
              (13.07.1;  
               13.25.3;  
               08.06.1;  
          )}

   \title{Search for GRB X-ray afterglows in the ROSAT all-sky survey}

   \author{J. Greiner\inst{1}, D.H. Hartmann\inst{2},
           W. Voges\inst{3}, 
           T. Boller\inst{3}, R. Schwarz\inst{1}, S.V. Zharikov\inst{4,5}}

   \offprints{J. Greiner, jgreiner@aip.de}

   \institute{Astrophysical Institute
        Potsdam, An der Sternwarte 16, 14482 Potsdam, Germany
       \and
        Clemson Univ., Dept. of Physics and Astronomy, Clemson, SC 29634, USA
       \and
        MPI for Extraterrestrial Physics , 85740 Garching, Germany
       \and
        Special Astrophysical Observatory, 357147 Nizhnij Arkhyz, Russia 
       \and
        Instituto de Astronomia, UNAM, 22860 Ensenada, Mexico
       }

   \date{Received 23 July 1999 / Accepted ??  October 1999}

   \authorrunning{Greiner \etal}

   \maketitle

\begin{abstract}
We describe a search for X-ray afterglows from gamma-ray bursts
using the ROSAT all-sky survey (RASS) data. If the emission
in the soft X-ray band is significantly less beamed than in the 
gamma-ray band, we expect to detect many afterglows in the RASS.
Our search procedure generated 23 afterglow candidates, where
about 4 detections are predicted. However, follow-up spectroscopy
of several counterpart candidates strongly suggests a flare star 
origin of the RASS events in many, if not all, cases. Given the 
small number of events we conclude that the ROSAT survey data are 
consistent with comparable beaming angles in the X-ray and gamma-ray
bands. This result
is perhaps not surprising, given that the data constrain the
relative beaming fraction only within a few hours of the burst. 
However, models predicting a large amount of energy emerging as a
nearly isotropic X-ray component of the early afterglow are severely
constrained by the ROSAT data. In particular, a so far undetected
class of ``dirty fireballs'' and delayed ``rebursts'' are constrained.

   \keywords{Gamma-rays: bursts --   X-rays: general --- Stars: flare
               }
\end{abstract}

\section{Introduction}

The discovery of fading X-ray afterglows from 
gamma-ray bursts (GRBs) with BeppoSAX (Costa {\it et al.} 1997;
Piro {\it et al.} 1998a) allowed the first identification
of these enigmatic events outside of the gamma-ray band. The 
subsequent discoveries of optical and radio afterglows associated 
with a large fraction of the fading X-ray sources led to the 
discovery of faint, extended host objects (``galaxies'', although this
identification is usually not well established). Spectroscopy 
resulted in the discovery of absorption lines in the otherwise
smooth continua of these optical transients (OTs). The corresponding
redshifts settled the previous debate of the GRB distance scale,
placing a typical GRB in the cosmological redshift range z = 1$-$2. 
At present, multi-wavelengths observations of GRB afterglows exist
for about 20 bursts (see http:/$\!$/www.aip.de/People/JGreiner/grbgen.html 
for a continuously
updated list of GRBs with afterglow emission, and Hartmann 1999 for
for a recent review of these exciting developments up to GRB 990123). 
These distances imply very large energies of the burst as well 
as their afterglows. Depending on uncertain beaming fractions 
GRBs require E$_{\rm grb} \sim 10^{52-54} {\rm erg}$, which constitutes
an unprecedented challenge to theorists. Geometric beaming might be
required to alleviate these energy requirements. The integrated afterglow
emission (all non-$\gamma$ bands) requires a comparable amount
of energy (with a large dynamic range in E$_{\rm grb}$/E$_{\rm agl}$).
This fact led to the descriptive term ``hypernova'' (Paczynski 1998)
for GRB afterglows whose light output can easily dominate other
explosive phenomena such as novae and supernovae.  
 
   \begin{figure*}
    \resizebox{12.5cm}{!}{\includegraphics{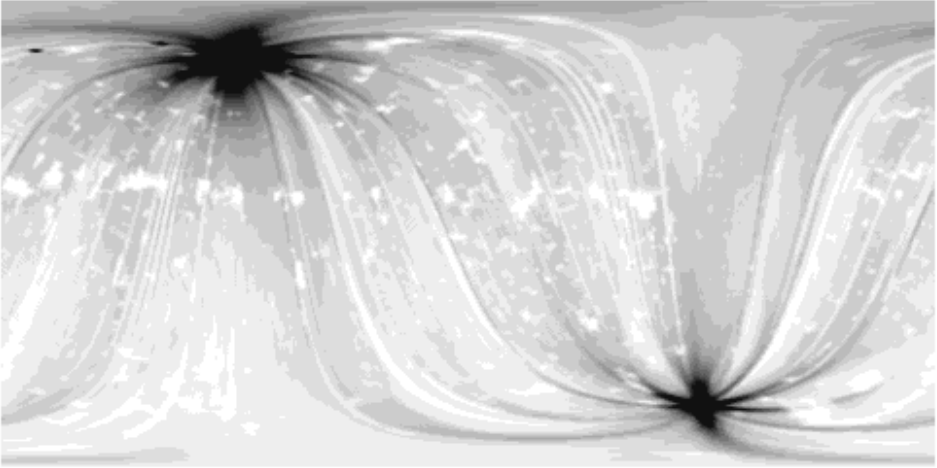}}
    \hfill
    \parbox[b]{50mm}{
    \caption[lc]{Exposure map of the ROSAT all-sky survey including 
        ``repairs'' in equatorial coordinates / rectangular projection
         with (0,0) in the middle. 
      The pattern is 
        due to the scans being performed in great circles at constant 
        ecliptic longitude. Therefore, the ecliptic poles (the two black 
        ``stars'') received highest exposure of
        up to 40 ksec (from Voges \etal\ 1999).}
        }
      \label{expomap}
   \end{figure*}

Deposition of such huge amounts of energy on a short time scale
into a small volume inevitably leads to the development of an opaque
electron/positron-photon fireball which quickly accelerates into 
the relativistic regime (e.g., Meszaros $\&$ Rees 1997; Piran 1999;
Meszaros 1999). The likely presence of baryons quenches immediate gamma-ray
emission, because by the time the expanding fireball becomes semi-
transparent to it's internal high energy photons most of the burst
energy is transferred to kinetic energy of the baryonic component
of the relativistic flow. In order to generate a GRB this energy
must be retrieved. This can occur through internal dissipation, 
which occurs when a central engine drives
multiple fireball shells with varying Lorentz factors such that
eventually shell-shell collisions lead to internal shocks which
dissipate energy via electron synchrotron radiation.
Alternatively, a GRB could occur when the relativistic shell(s)
interact with an external medium (ISM or matter ejected by the
progenitor prior to the burst). Consideration of the observed
highly structured GRB lightcurves suggests that the GRB itself
is more likely the result of internal shocks (Fenimore \etal\ 1996,
1999; Sari $\&$ Piran 1997; Kobayashi, Piran, $\&$ Sari 1997; but 
see Dermer $\&$ Mitman 1999)
while the external shocks are thought to be responsible
for the smoother afterglow emission (e.g., Piran 1999). However,
it is far from clear whether or not the observed X-ray afterglows
are exclusively due to external shocks. Some overlap with the prompt
emission from internal shocks 
is conceivable. If afterglow and burst emission are from separate 
regions one must seriously consider the possibility that prompt 
$\gamma$-ray and delayed X-ray emission are beamed (if at all) differently.
If so, one expects X-ray afterglows to be less beamed than GRBs. 
This possibility can be tested with independent searches for 
afterglows in existing X-ray surveys (e.g., Grindlay 1999). Wavelengths
bands other than X-rays also offer potential means to constrain 
differential beaming (Meszaros, Rees, \& Wijers 1999) through
supernova searches in the optical band (Rhoads 1997), searches
for radio through afterglows (Perna \& Loeb 1998), or dedicated GRB/OT 
surveys such as LOTIS (Park {\it et al.} 1997),
ROTSE (Akerlof {\it et al.} 1999), TAROT (Boer {\it et al.} 1998), 
and similar programs under development worldwide.

Although we are far from a complete theoretical understanding of
the various emission components of a burst (prompt gamma-ray to
optical emission, delayed afterglow emission of power-law form
with slope changes at late times (also evidence for beaming), and
the appearance of ``Supernova emission'' at late times) it is 
a straightforward observational task to check whether or not possibly
wavelength-dependent beaming exists. 
Popular GRB scenarios such as binary coalescence of compact stars (e.g., Janka $\&$
Ruffert 1996) or collapsars (Woosley 1993; MacFadyen $\&$ Woosley 1999;
Hartmann $\&$ MacFadyen 1999) predict strongly collimated flows, which should 
also lead to strongly collimated burst and afterglow emission. If afterglows 
turn out to be less beamed {\it relative} to the GRBs, then we expect to find a 
higher rate of afterglows than GRBs. We test this possibility with
a search for X-ray afterglows that were fortuitously detected during
the ROSAT all-sky survey. Preliminary results of this study were
reported by Greiner \etal\ (1999).

\section{ROSAT all-sky survey data and expectations for transient
afterglow detections}

The ROSAT satellite (Tr\"umper 1983) performed the first 
all-sky survey in the 0.1--2.4 keV
X-ray band during 1990 August 1  -- 1991 January 25 with short additional
exposures (``repairs'') carried out in February (16--18) and 
August (4--12) of 1991 (Voges, {\it et al.} 1999). 
During the satellite's orbital period of 96 minutes the telescope (with a
field of view diameter of 2\degr) scans a full 360\degr\ circle
on the sky. Thus, the exposure (per scan) for a source located inside the 
scan circle is typically in the range
10--30 sec. Due to orbital plane rotation (together with Earth's
motion) these full circles move with 1\degr/day perpendicular to the
scan direction, covering the whole sky in 6 months. Thus, a source located 
near the ecliptic equator is covered by the telescope scans during a
period of about two days. However, this coverage 
increases to 180 days at the ecliptic poles. Sky exposure is thus a 
very sensitive function of ecliptic latitude, with typical exposures of
$\sim$ 400 sec near the equator and up to 40 ksec very close to the poles.
Figure \ref{expomap} shows the exposure map of the RASS, but we note that our
study relies on the product of exposure in time and coverage in area
so that the large exposure at the poles and low equatorial exposure
is compensated by the correspondingly small/large solid angles (according
to $\cos$(ecliptic latitude)). The 
net effect is a rather uniform search pattern. 

Even with a single exposure of 10--30 s duration
the sensitivity of ROSAT in the 0.1--2.4 keV band is sufficient to detect 
GRB X-ray afterglows for several hours after the burst. 
Fig. \ref{rossen} shows this single-scan sensitivity of the ROSAT
PSPC in comparison to several recent X-ray afterglow light curves observed
by BeppoSAX. 
The fraction, $f$, of afterglows detectable during the RASS depends 
critically on three parameters: 

   \begin{figure*}
   \vspace*{0.2cm}
    \resizebox{\hsize}{!}{\includegraphics{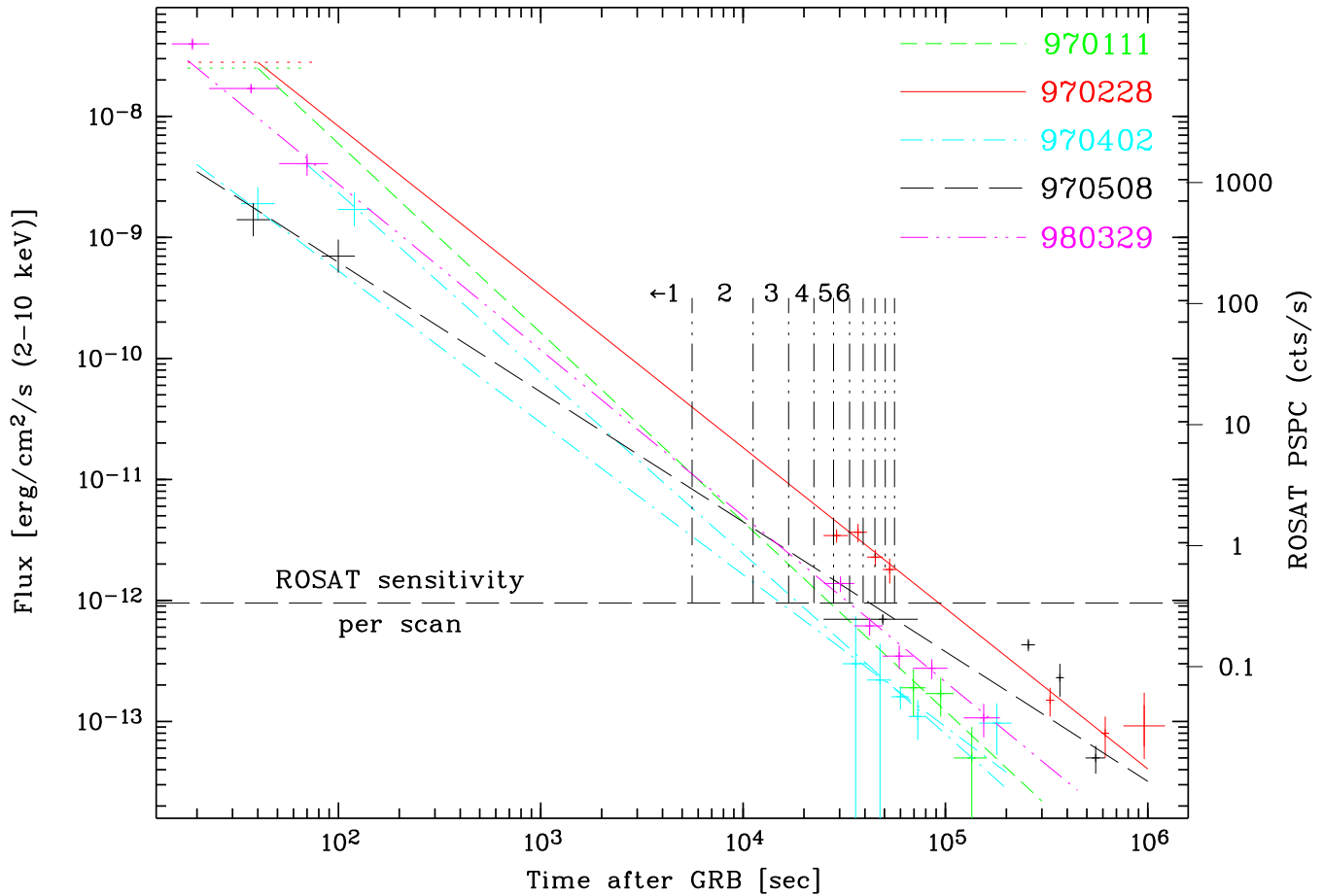}}
    \vspace*{-0.7cm}
    \caption[lc]{Afterglow light curves of some observed GRB X-ray 
        afterglows in the 2--10 keV range 
        (GRB 970111: Feroci \etal\ 1998; 
         GRB 970228: Costa \etal\ 1997;
         GRB 970402: Nicastro \etal\ 1998;
         GRB 970508: Piro \etal\ 1998b; 
         GRB 980329: in 't Zand \etal\ 1998)
        and their corresponding brightness extrapolated
        into the ROSAT band (scale on the right; assuming a power 
        law with photon index of --2 and neglecting foreground
        absorption). The horizontal line gives the sensitivity
        of the ROSAT PSPC during one scan, and the vertical lines mark
        the time windows for the possible coverage of a GRB location
        by ROSAT during its scanning mode. Thus, one anticipates afterglow
        intensities of several hundred cts/s during the first
        scan, a few to ten cts/s during the second scan, 
        less than 2 cts/s during the third scan, and so on.}
      \label{rossen}
   \end{figure*}

The first contributing factor is the 
fraction of GRBs that have detectable X-ray afterglows. 
Observations with BeppoSAX
indicate that this fraction is rather close to one. In addition, the 
burst monitor of BeppoSAX appears to sample the full flux or
fluence range observed by the BATSE detectors. SAX does not select
against faint bursts. However, the trigger algorithm of SAX has
in fact introduced a bias towards bursts of long durations (exceeding
about one second), so that it is currently impossible to argue about
X-ray afterglows from short bursts, which might be due to mergers of
neutron star binary systems (e.g., Janka $\&$ Ruffert 1996). It is 
conceivable that these events have drastically different X-ray (and
other) afterglows with correspondingly different beaming behavior. At
this point in time we are not in the position to draw conclusions
from the RASS about such events. The proposed SWIFT mission and the 
forthcoming HETE2 mission might rectify this situation in the near
future. 

   \begin{figure*}
    \vspace*{0.2cm}
    \resizebox{12.5cm}{!}{\includegraphics{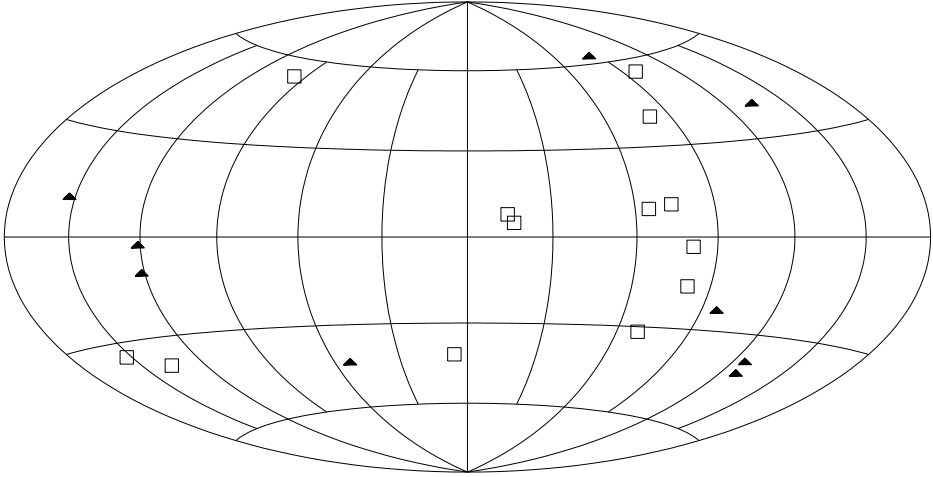}}
    \hfill
    \parbox[b]{50mm}{
    \caption[gal]{Distribution of the selected 23 afterglow candidates
        in galactic coordinates (galactic center is in the middle).
        Single peak events (SP in Tab. \ref{xcand}) are shown as filled
        triangles, while the declining events are shown as open 
        squares. This distribution is biased by the strong anisotropy of the
        exposure (see Fig. \ref{expomap}; but note the different 
        coordinate system). }
        }
      \label{galdis}
   \end{figure*}

The second relevant factor determining the expected event rate
is a possible correlation of X-ray flux to $\gamma$-ray peak flux (or
fluence, or some other characteristic aspect of the GRB itself). So far, 
the observed X-ray afterglow fluxes measured about 100 sec after
the GRB are spread within a factor of 10, while the GRB fluxes show 
a dynamic range in excess of 1000. The fluence range is also much
larger. It is thus impossible to accurately predict the properties 
of the X-ray afterglows based on direct GRB observations. On the 
other hand, the small dispersion of X-ray afterglows provides some
confidence that we can use a mean X-ray afterglow template for the
purpose of this study. We are not likely to introduce a strong bias
against certain types of X-ray afterglows. However, one should be
aware of the possibility of strong selection against rapidly decaying
afterglows (see below) and the fact that little information exists on 
afterglows from short burst (see discussion above). For the purpose of 
this study we assume that present afterglow data provide a
representative sample.

The third factor is the X-ray intensity decay law: SAX observations
have firmly established that the typical afterglow is a power law 
t$^{-\alpha}$ (although bumps and wiggles exist in several bursts) 
with values of the index ranging from $ -2 < \alpha < -1$.

\begin{table*}
\caption{X-ray afterglow candidates selected from the RASS:
  Given for each X-ray source are the RASS source name (column 1),
  the statistical positional error (2), the total number of counts (3),
  the detection likelihood ML (4), the total exposure time during the RASS (5),
  the maximum count rate (6), the number of scans over the source (7),
  two measures for the amplitude and significance of the variability (8, 9),
  a flag indicating single-peaked light curves (10) and comments on optical
  data (11). } 
\vspace{-0.25cm}
\begin{tabular}{ccccccccrcc}
\noalign{\smallskip}
\hline
\noalign{\smallskip}
 Source Name & Error & cts & ML$^1$ & T$_{exp}$ & CR$_{max}$ & Scans & 
    S/N$^2$ &    VI$^3$ & SP$^4$ & Comments \\
             & (\asec)   &    &    & (sec)    & (cts/s) &         &     & 
                  &  &           \\
\noalign{\smallskip}
\hline
\noalign{\smallskip}
1RXS J003528.6+603139 &  9 &  30 & 47 & 349 & 1.1 & 17  & 3.2 &  3.6 & sp & \\
1RXS J004031.1+520906 &  7 & 120~ & 373 & 453 & 5.4 & 23  & 9.1 & 23.6 & sp & \\
1RXS J013556.7+231605 & 16 &  31 & 38 & 441 & 1.7 & 19  & 3.9 &  6.6 &    & K(?) star (Hamburg)\\
1RXS J023644.4+224028  & 10 &  27 & 63 & 247 & 0.9 & 15  & 3.0 &  3.3 &    & \\
1RXS J043412.3--314911 &  8 &  37 & 83 & 258 & 1.3 & 15  & 3.9 &  4.7 & sp & \\
1RXS J045248.0--324507 & 12 &  17 & 27 & 286 & 1.3 & 17  & 3.5 &  5.5 & sp & \\
1RXS J050154.6--785616 &  9 & 176~ & 357 & 630 & 5.5 & 33  & 8.0 & 14.8 &    & dMe (RBS) \\
1RXS J051515.9+574705  & 10 &  25 & 60 & 406 & 1.0 & 19  & 3.3 &  4.4 & sp & \\
1RXS J061909.0+083859  &  7 &  25 & 47 & 473 & 3.3 & 25  & 6.7 &  7.7 & sp & M3.5e (spectrum) \\
1RXS J064118.6--543503 & 23 &  33 & 30 & 651 & 3.6 & 29  & 4.4 & 14.2 & sp & M4.5e (spectrum) \\
1RXS J081727.0--650718 & 12 &  49 & 28 & 964 & 1.0 & 47  & 3.5 &  5.0 &    & M5e (spectrum) \\
1RXS J093800.6+081640 & 10 &  20 & 39 & 210 & 1.5 &  ~9  & 3.5 &  4.8 & sp & M3.5e (spectrum) \\
1RXS J094037.3--565615 & 16 &  24 & 37 & 299 & 1.0 & 19  & 3.1 &  4.5 &    & \\
1RXS J111055.7--485510 & 17 &  38 & 65 & 135 & 2.1 & ~5  & 3.9 &  7.4 &    & \\
1RXS J112511.7--002437 & 15 &  41 & 48 & 428 & 2.3 & 23  & 4.1 &  6.2 &    & M2.5e (spectrum) \\
1RXS J113523.0--191321 & 19 &  32 & 37 & 341 & 1.1 & 19  & 3.0 &  3.0 &    & M4.5e (spectrum) \\
1RXS J115928.5--524717 & 11 &  27 & 56 & 266 & 1.0 & 13  & 3.2 &  4.1 &    & \\
1RXS J120328.8+024912 & 19 &  61 & 66 & 404 & 3.6 & 21  & 6.1 & 15.6 & sp & empty on DSS2 \\
1RXS J144713.2+570205 &  8 &  87 & 132 & 746 & 2.1 & 41  & 4.5 &  6.3 &    & M star (Hamburg) \\
1RXS J163607.8--354353 &  8 &  36 & 36 & 316 & 1.2 & 15  & 3.3 &  3.8 &    & \\
1RXS J163947.8--392023 & 17 &  63 & 99 & 336 & 1.9 & 17  & 3.2 &  2.7 &    & \\
1RXS J210246.3--372149 &  8 &  78 & 225 & 306 & 1.5 & 17  & 3.4 &  3.1 & & dMe (RBS) \\
1RXS J215651.5--050608 & 33 &  22 & 15 & 263 & 1.6 & 19  & 3.2 &  4.8 & sp & \\
\noalign{\smallskip}
\hline
\end{tabular}
\label{xcand}

\noindent{\small
      $^1$ Maximum likelihood of the source detection,
           defined as --ln(P), where P is the probability that
           the observed distribution of photons originates from 
           a spurious background fluctuation. \\
      $^2$ The signal-to-noise ratio S/N is defined as the ratio of
           maximum count rate minus the mean count rate outside the maximum
           over the square root of the quadratic sum of their errors. \\
      $^3$ The variability index VI is defined as the ratio of
           maximum count rate minus its error over the mean count rate 
           outside the maximum plus its error.     \\
      $^4$ Light curves displaying only one bin with non-zero count rate
           are labeled as ``single-peaked'' (sp). 
         }
\end{table*}

It is currently not clear how one should combine all these factors
into a proper statistical distribution from which to derive the overall
sampling fraction $f$. We thus simply use the existing database
as a representative set of templates and compare this set to the 
ROSAT PSPC sensitivity. For this comparison we estimate the flux in
the ROSAT PSPC from an extrapolation of the flux measured by SAX assuming 
a power law photon spectrum with a universal slope $-$2. Possible 
foreground absorption was neglected; the fraction of the full sky
for which the effective hydrogen column density is large enough to
remove afterglows completely is about 2$\%$. The resulting comparison
implies that the RASS would in fact be sensitive enough to
detect all GRB afterglows in 3 subsequent scans, and $\sim$80\% 
in 5 scans (see Fig. \ref{rossen}). We adopt a conservative fraction
of $f=0.8$ for the subsequent analysis.

The above comparison and Fig. \ref{rossen} show that GRB afterglows 
have been detectable during the RASS for about 1-5 scans, thus implying that
the RASS data sample a GRB afterglow light curve at a time of $\sim$1--8 hrs
after the GRB. We note that this time span so far is completely unstudied.

The number of detectable X-ray afterglows from GRBs beamed towards us 
(based on the BATSE detection rate) during the RASS is
$$ N^{agl} = f \times S_R^{agl} \times R_{GRB} \ \ \ , $$
where $R_{GRB}$ is the rate density of GRBs (bursts per unit time and
unit solid angle)
and $S_R^{agl}$ is the RASS afterglow coverage function in units of
time$\times$area. We adopt 
$R_{GRB}$ = 900 GRBs/sky/yr $\equiv$ 1 GRB/(16628$\Box \degs \times$ days).
$S_R^{agl}$ would be 122296.5 $\Box \degs \times$ days for 100\% coverage 
in time. The temporal completeness of the RASS was 62.5\% (Voges \etal\ 1999),
so that $S_R^{agl}$ = 76435  $\Box \degs \times$ days.
Thus, we expect N$^{agl}$ = 4.6$\times$$f$ $\sim$ 3.7 GRB afterglows 
to be detected during the RASS. 

We note here that the coverage function of ROSAT is very different for 
prompt GRB emission (with duration of seconds) and X-ray afterglows 
(with duration of hours).
$S_R^{agl}$ would have to be replaced by $S_R^{GRB}$ corresponding to
the mean exposure per sky location times the full sky:
$S_R^{GRB}$ = 359 $\Box \degs$, i.e. a factor of 360 lower than $S_R^{agl}$.
The ROSAT survey is thus too limited for meaningful constraints on beaming 
patterns 
in prompt X-ray emission from GRBs. Woods $\&$ Loeb (1999) used the {\it Ariel
V} catalog of fast transients (Pye $\&$ McHardy 1983) to place constraints
on beaming during the burst, but concluded that {\it Ariel}'s sensitivity
is not great enough. While beaming during the GRB is thus the domain of future
detectors, constraints on long-duration afterglows can be achieved with
existing surveys, such as the RASS.

\section{The search for afterglow candidates}

\subsection{ROSAT X-ray data}

We first produced scan-to-scan light curves for all RASS sources
with either a count rate larger than 0.05 cts/s or a detection likelihood
exceeding 10, resulting in a total of 25,176 light curves. 
Note that these criteria correspond to a lower sensitivity
threshold in comparison to the RASS Bright Source Catalog which used a 
minimum of 15 counts and a detection likelihood $\gax$ 15 (Voges \etal\ 1999).
Each of these light curves consists of about 20 to 450 bins spaced at 96 min.,
with each bin corresponding to 10--30 seconds exposure time.

After ignoring 363 light curves with negative mean count rates (caused
by incorrect background-subtraction in the automatic procedure) 
we apply three selection criteria to these light curves:

(1) The maximum bin should have a signal-to-noise ratio of S/N$>$3 above the 
mean count rate around the maximum. S/N is defined as the difference 
between the
maximum and mean count rate divided by the square root of the quadratic
sum of the error of the maximum and mean count rates.
Note that this S/N ratio is a measure of the variability amplitude,
but not of the significance of the peak or the X-ray source itself
(see column 4 of Tab. \ref{xcand} for this latter significance).

(2) The mean count rate derived from observations obtained until one bin 
prior to the maximum count rate should be consistent with zero. This
criterion allows a transient to rise within the width of one bin 
($\sim$1.5 hrs), but not slower than that. 

(3) Similar to condition (2) we demand that the mean count rate at 
times later than those covered by 5 bins past maximum should also be
consistent with zero. Like the previous condition, this requirement 
suppresses transient sources that have quiescent emission at detectable 
levels, such as nearby flare stars. In fact, when we do not require this
condition a significant set of well known (and new) flare stars appears.

Application of the above listed criteria yields a total of 32 GRB
afterglow candidates.
We then proceed with additional conditions that proper afterglows
should display:

(i) Sources with double and multipeak structures are excluded, simply
because this pattern does not fit that of ``standard'' X-ray afterglows
from GRBs (four transients removed).

(ii) Sources with a rise extending over several bins and showing zero 
flux immediately after the peak (inverse afterglow behavior) are also
selected out for obvious reasons (this removes two transient sources).

(iii) We also investigated pointed ROSAT observations, which were available for
3 of the remaining candidates. Two sources were found to be unacceptable
candidates because they did exhibit persistent 
X-ray emission at a level below the RASS threshold.

   \begin{figure*}
    \vbox{\psfig{figure=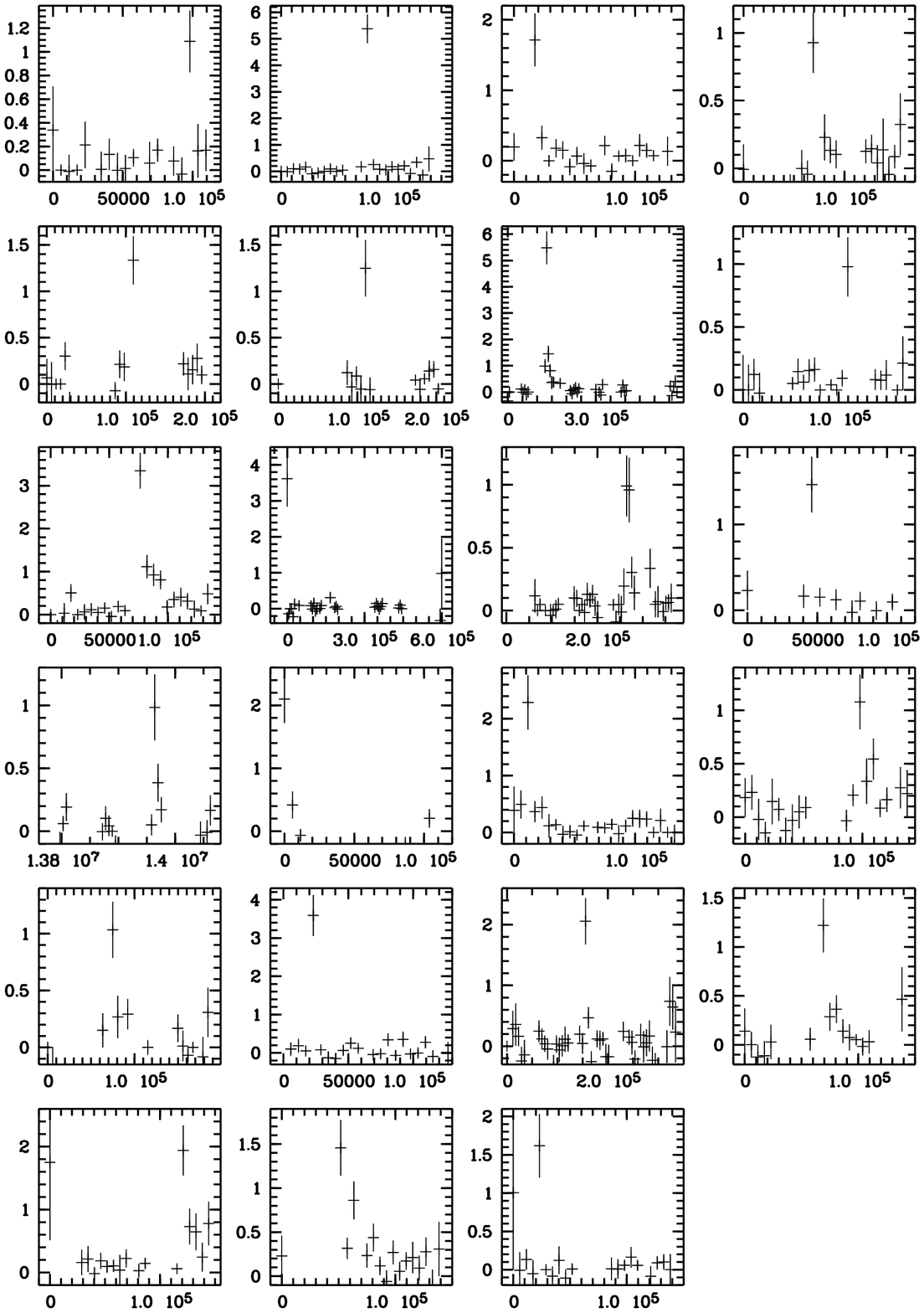,width=16.cm,%
           bbllx=2.3cm,bblly=3.1cm,bburx=19.2cm,bbury=26.8cm,clip=}}\par
    \vspace*{-.1cm}
    \caption[lc]{RASS X-ray lightcurves of all 23 sources listed in 
         Table \ref{xcand} in identical order as that of Fig. \ref{find}.
         Units are ROSAT PSPC counts/sec for the y-axis, and time in seconds
         for the x-axis. Time zero corresponds to the first scan of the
         PSPC field-of-view over the source. }
      \label{lc}
   \end{figure*}

   \begin{figure*}
    \vbox{\psfig{figure=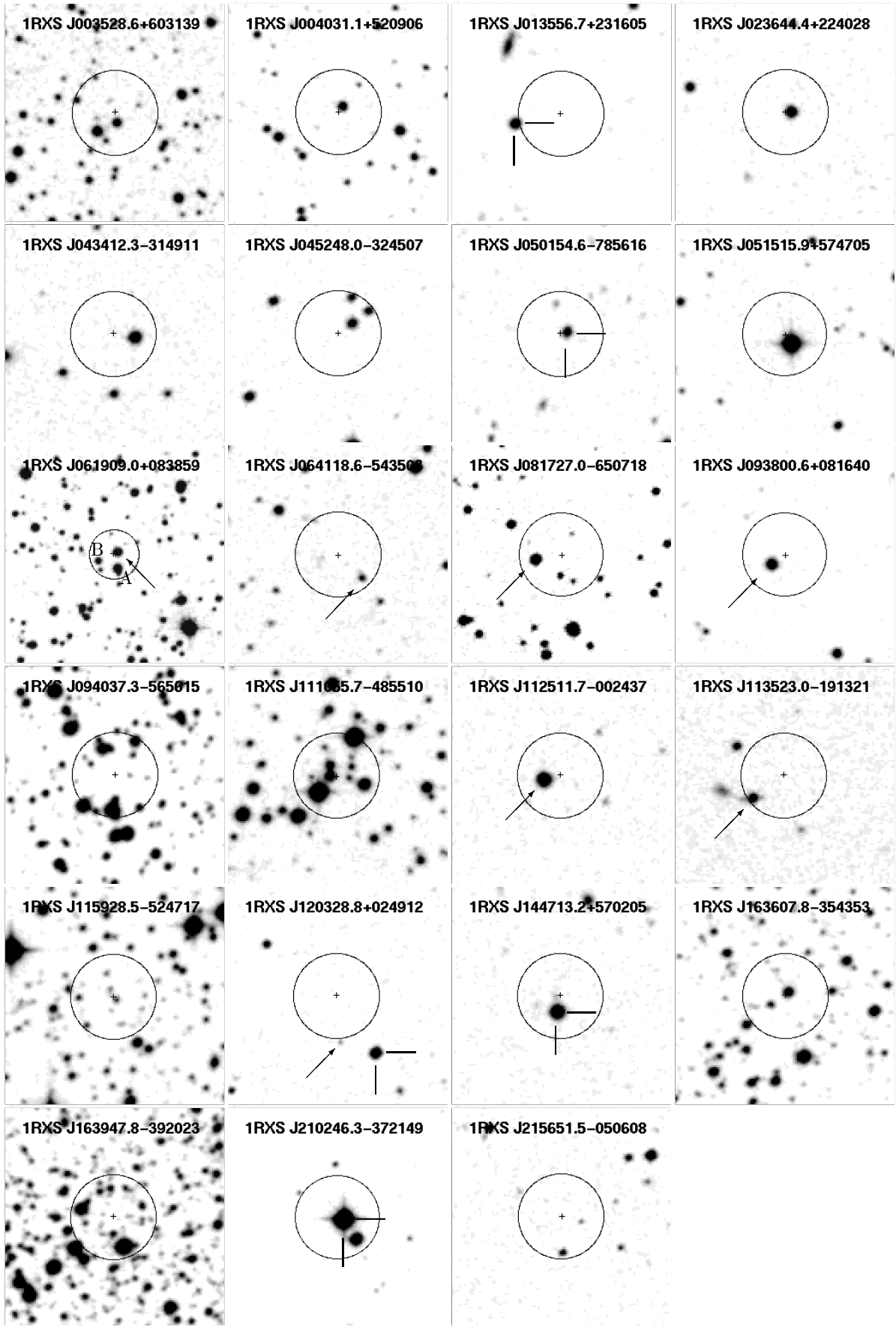,width=16.cm,%
       bbllx=2.0cm,bblly=0.8cm,bburx=18.7cm,bbury=25.7cm,clip=}}\par
    \vspace{-0.2cm}
    \caption[fc]{DSS finding charts of all candidates listed in
       Table \ref{xcand}. The 90$\%$ confidence error circles have identical
       radii of 25\asec\ (based on the ROSAT Bright Survey Catalog statistics; 
       Voges \etal\ 1999). 
       Table \ref{xcand} provides individual, 
       statistical RASS source localisation errors. 
       For the objects depicted by arrows
       optical spectra have been obtained (see Fig. \ref{ospec6}),
       and for those  marked by two dashes Hamburg objective prism or RBS
       identifications are available. }
      \label{find}
   \end{figure*}

\begin{table*}
\caption{Details on proven and suspected flare stars within the sample of 23
  X-ray afterglow candidates}
\vspace{-0.3cm}
\begin{tabular}{cccccccc}
\noalign{\smallskip}
\hline
\noalign{\smallskip}
 Source Name & optical coordinates & r$_{USNO}$  & (b-r)$_{USNO}$ 
      & V$_{spectrum}$ & spectral & distance & L$_{\rm X}^{\rm peak}$ \\
             &  (2000.0)$^{(1)}$       & (mag)       & (mag)  
      & (mag) & type     & (pc)     &   (10$^{31}$ erg/s)$^{(2)}$ \\

\noalign{\smallskip}
\hline
\noalign{\smallskip}
 1RXS J004031.1+520906  & 00 40 30.8  +52 09 10 & 15.5  & 2.7 & & & & \\
 1RXS J013556.7+231605  & 01 35 58.6  +23 15 58 & 15.2  & 3.1 & & K(?) & & \\
 1RXS J023644.4+224028  & 02 36 44.1  +22 40 29 & 13.6  & 3.9 & & & & \\
 1RXS J043412.3--314911 & 04 34 11.3 --31 49 13 & 14.2  & 2.5 & & & & \\
 1RXS J050154.6--785616 & 05 01 51.9 --78 56 17 & --    & -- & & dMe & & \\
 1RXS J051515.9+574705  & 05 15 15.4  +57 46 59 & 11.0  & 0.9 & & & & \\
 1RXS J061909.0+083859  & 06 19 09.0  +08 39 03 & 13.7  & 2.7     
      & 14.7 & M3.5e & 40 & 1.5 \\
 1RXS J064118.6--543503 & 06 41 17.0 --54 35 18 & 18.4  & 3.4 
      & 19.2 & M4.5e & 250~\, & 63.0~\, \\
 1RXS J081727.0--650718 & 08 17 29.5 --65 07 21 & 16.9  & 2.6 
      & 17.2 & M5.0e & 70 & 1.8 \\
 1RXS J093800.6+081640  & 09 38 01.1  +08 16 34 & 13.8  & 3.2 
      & 15.4 & M3.5e & 70 & 1.5 \\
 1RXS J112511.7--002437 & 11 25 12.4 --00 24 38 & 13.3  & 2.8 
      & 14.3 & M2.5e & 60 & 1.7 \\
 1RXS J113523.0--191321 & 11 35 24.9 --19 13 34 & 15.1  & 3.2 
      & 16.8 & M4.5e & 90 & 2.0 \\
 1RXS J144713.2+570205  & 14 47 13.5  +57 01 55 & 13.2  & 2.6 &  & M & & \\
 1RXS J210246.3--372149 & 21 02 46.0 --37 21 51 & -- & $\gax$1.7~ & & dMe & &  \\
 1RXS J215651.5--050608 & 21 56 51.4 --05 06 30 & 17.5  & 1.5 & & & & \\
\noalign{\smallskip}
\hline
\end{tabular}
\label{sptype}

\noindent{\small
       $^{(1)}$ Coordinates have been measured on the DSS2, and thus have
            a mean error of $\pm$1\asec. \\
       $^{(2)}$ The luminosities have been determined in the 0.1--2.4 keV range
            under the assumption of a 1 keV thermal bremsstrahlung spectrum.
            The bolometric luminosity of such a spectral model is a factor
            1.3 larger than that given for the 0.1--2.4 keV range. 
         }
\end{table*}

(iv) Finally, we correlated the candidate list with various optical, 
infrared and radio catalogs, and excluded one X-ray source which has
a 9th magnitude, seemingly active star (HD 101082) in its error circle.

The application of these selection steps yields a total of 23 transients
as viable X-ray afterglow candidates. Table 1 summarises the relevant
properties of these events including the significance of the sources 
(column 4), measures of the amplitude and signigicance of the transient
behaviour (cols. 8, 9). Figure \ref{lc} shows the individual light curves
and Figure \ref{find} provides DSS images of the X-ray positions. The 
interpretation of these data is given in the next section.

Inspection of the candidate list presented in Tab. \ref{xcand} and 
Fig. \ref{lc} shows that about 50$\%$ of these light curves display single 
peaks, i.e. outbursts with just one bin satisfying S/N$>$3 and otherwise zero 
count rate. The remainder shows decays that more closely resemble 
GRB afterglow behavior.

Many of the events in the table (single
peak SP, or declining) could be flare stars (the SP sources might
also have a significant fraction of statistical fluctuations), but
an identification of these events as stellar 
flares requires optical follow-up studies. The durations of the single
bin events are consistent with time scales of flares from late-type stars
(10--60 minutes), but even the declining events do not have an unreasonably 
long duration. 
Also, the distribution on the sky does not reveal any systematic difference
between single peak events and the rest (Fig. \ref{galdis}).

\subsection{Optical data}

\begin{figure*}
    \vbox{\psfig{figure=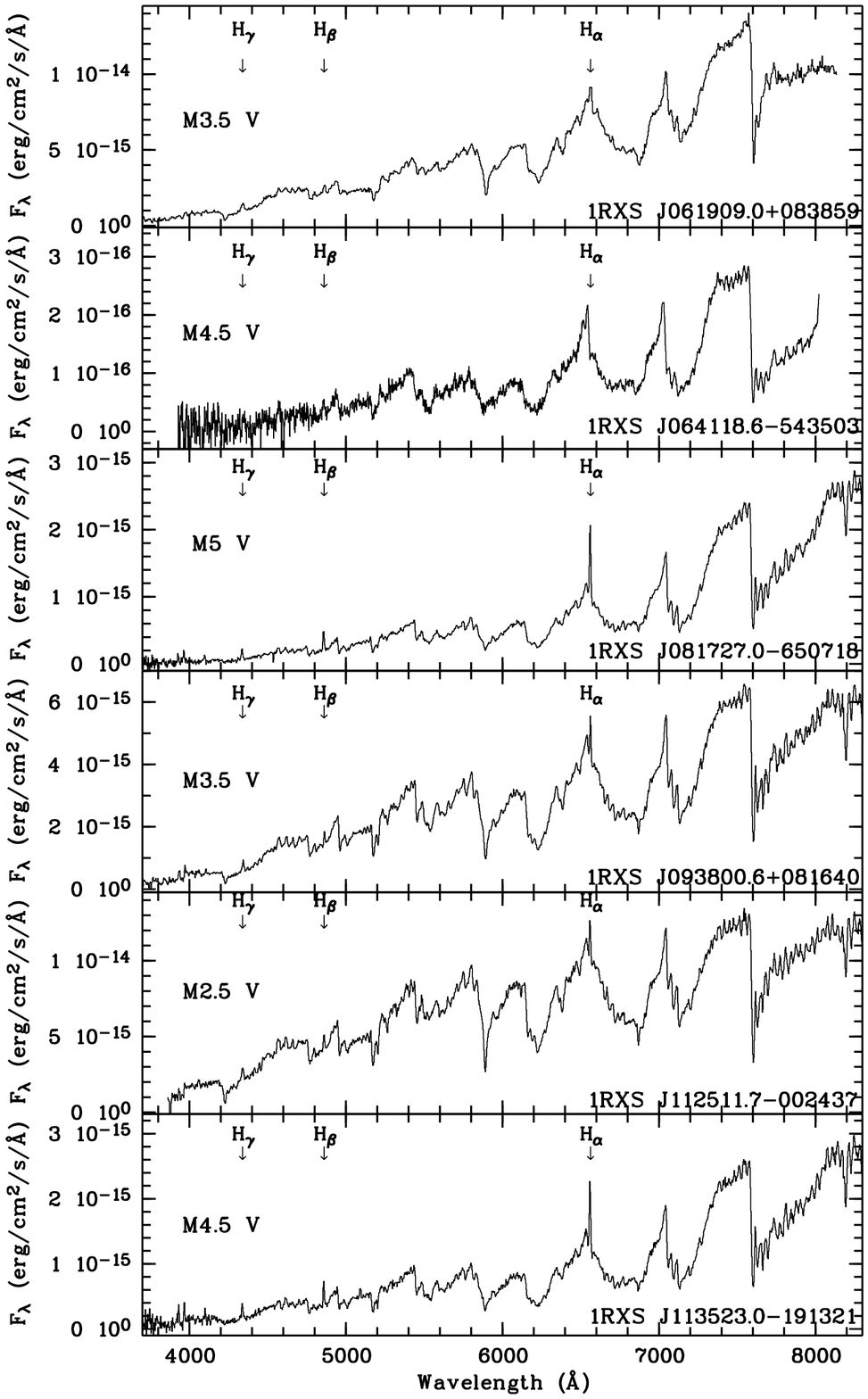,width=14.1cm,%
           bbllx=1.5cm,bblly=1.2cm,bburx=17.2cm,bbury=26.4cm,clip=}}\par
    \caption[ospec]{Spectra of six randomly selected counterpart
        candidates. In all cases the spectrum refers to the 
        object indicated with an arrow in Fig. \ref{find}.}
      \label{ospec6}
   \end{figure*}

To estimate the flare star fraction of the events listed in 
Table \ref{xcand} we obtained optical spectra for
six randomly selected bright sources inside the X-ray error circles. 
Three different telescopes were used to acquire these spectra:
the 6m telescope of SAO (December 17, 1998; top panel of Fig. \ref{ospec6}), 
equipped with the spectrograph in the Nasmyth-1 focus,
the 3.6m telescope at La Silla/ESO (January 23, 1999; second panel
of Fig. \ref{ospec6}) equipped with EFOSC,
 and the Danish 1.5m telescope at La Silla/ESO (January 26, 1999;
remaining 4 panels of Fig. \ref{ospec6}) equipped with DFOSC.
Grisms with 250, 300 and 300 grooves per mm was used yielding
a dispersion of 300 \AA/mm, 140 and 220 \AA/mm, respectively. 
With a 2\asec, 1\farcs5 and 1\farcs5 slit 
the FWHM resolution is 16 \AA, 14 \AA\ and 12 \AA, respectively. 
Exposure times range from 600--1200 sec, and the spectra were
debiased, flatfielded and calibrated (with the standard star G191B2B or GD 108)
using standard MIDAS procedures.  Telescope time
constraints did not allow us to obtain more spectra than those indicated
in Fig. \ref{find}.

All six objects which are the brightest objects within the respective X-ray 
error circle, turn out to be M stars with strong emission lines of the Balmer
series (Fig. \ref{ospec6}). Given the fact that in five cases these Me stars
are the only optical object down to the POSS limit, it is secure to
identify the corresponding X-ray sources as being due to X-ray flares from
these Me stars. For the sixth object, 1RXS 061909.9+083859, two other
bright stars are inside the X-ray error box (denoted ``A'' and ``B'' in
Fig. \ref{find}). Their spectra, however,
indicate F/G spectral types, and thus (based on the 
$L_{\rm X}/L_{\rm opt}$ ratio) argue against one of these being the
optical counterpart of the X-ray source.
Based on the optical brightness of the six flare stars
and the well-known $L_{\rm X}/L_{\rm opt}$ ratio of 1/50...1/100 
the expected X-ray intensity during quiescence is 
1$\times$10$^{-14}$...2$\times$10$^{-13}$ erg/cm$^2$/s.  This corresponds
to ROSAT PSPC count rates of 0.0015...0.03 cts/s and is below the RASS 
sensitivity, thus consistent with the non-detection outside the X-ray flare.

We have attempted a crude spectral classification of these six Me stars
to gain a little more inside into their properties. Following the method
initiated by Young \& Schneider (1981) and Wade \& Horne (1988)
and further developed in Schwarz \etal\ (1998) we used the strength of the
TiO bands to determine spectral classes. In particular, we determined
the continuum level outside the TiO bands, and then determined the flux 
deficits in the wavelengths
bands 6200--6220 \AA, 6760--6810 \AA, 7120--7150 \AA\ and 7650--7690 \AA.
A comparison of various ratios of these flux deficits with that of
well-known Gliese stars (Schwarz \etal\ 1998) results in the spectral types
listed in Tab. \ref{sptype}. Given the spectral resolution and the 
systematic errors in both, the flux calibration of the spectra as well
as the correlation of the TiO band ratios to spectral type we estimate
the error in our spectral class determination to be $\pm$1.

Based on these spectral types and the absolute magnitudes of M stars,
and allowing for extinction between 0.1-0.3 mag (corresponding to
half the total galactic extinction in the directions of the six Me stars)
we have derived a rough distance estimate for each star. Finally, we
converted the maximum X-ray count rate during the peak in the light curve
(Fig. \ref{lc} and Tab. \ref{xcand}) into an X-ray flux under the assumption 
of a 1 keV thermal
bremsstrahlung spectrum and extinction as above. Together with the
distances the derived peak X-ray luminosities during the corresponding
flare are also given in Tab. \ref{sptype}. These luminosities are typical
for flare stars.

In addition we also checked the Hamburg/RASS catalogue of optical 
identifications (Bade \etal\ 1998). These identifications are based
on a correlation of (an early version of) the ROSAT survey bright source
catalog with the data obtained with the Hamburg Schmidt telescope on
Calar Alto (Spain) during an objective prism plate survey of the northern
hemisphere.
Three out of our 23 objects were found
in this catalog:
\begin{itemize}
\vspace*{-0.2cm}
\item {\it 1RXS J013556.7+231605 $\equiv$ RX J0135.9+2316}:
The bright optical object on the border of the error circle 
(see Fig. \ref{find}) is identified as a star of spectral type K.
Due to the spectral resolution of less than 50 \AA\ it seems possible that
this star could also be of an early M spectral class, and thus
would possibly be a good flare star candidate.
\item {\it 1RXS J120328.8+024912 $\equiv$ RX J1203.4+0249}:
The bright optical object south-west of the X-ray error circle is
identified as a star of spectral type F/G. Therefore, a M flare star
origin is excluded, which anyway would have required the assumption of
a particularly bad X-ray position (it is located 45\asec\ off the centroid
X-ray position). Together with our optical identification of the
faint object just outside the X-ray error circle (being of F to G spectral 
type also and thus not a possible X-ray source counterpart and thus not
shown in Fig. \ref{ospec6}) we therefore 
conclude that the location of 1RXS J120328.8+024912 and its surrounding up to
nearly 1 arcmin is empty down to the limit of DSS2. 
\item {\it 1RXS J144713.2+570205 $\equiv$ RX J1447.2+5702}:
The bright and only optical object inside the error circle is
identified as a star of spectral type M, suggesting that this
could be a flare star as well.
\end{itemize}

Finally, two sources of our sample were already identified in the 
RBS programme (Schwope \etal\ 1999), a complete identification programme of
all bright ($>$0.2 cts/s), high-galactic latitude ($|b| > 30\degr$) sources   
in the RASS.
\begin{itemize}
\vspace*{-0.2cm}
\item {\it 1RXS J050154.6--785616}: The bright optical star inside the error 
  box is classified as dMe, so again is a flare  star.
\item {\it 1RXS J210246.3--372149}: The brighter of the two optical stars 
  is an early M type star with weak H$\alpha$ emission, and the
  fainter one a F/G type star. Thus, the dMe star is the most probable
  optical counterpart.
\end{itemize}

Inspection of the DSS finding charts (Fig. \ref{find})
suggests that all (but one) source have likely stellar counterparts.
There always (except for 1RXS J120328.8+024912) appears to be at least one
star in the RASS error box which could be the flare star responsible
for the spike detected in the RASS. In addition, all these stars have
very red colors as deduced from the USNO A1.0 catalog (Monet \etal\ 1998; 
see Tab. \ref{sptype}) supporting the conjecture that these are indeed
M stars.

\section{Discussion and Conclusions}

We thus argue that the bulk of the ``afterglows'' listed
in Table 2 are probably due to X-ray flares from nearby late-type stars.
It is of course impossible to rigorously prove this assertion until
spectroscopy has been obtained for all counterpart candidates. In the
meantime we argue that the existing data support the notion that the
RASS contains at most a few X-ray afterglows from GRBs.
This interpretation is consistent with the expected
number of afterglows (N$_{agl}$ = 3.7) derived in $\S$ 2.
1RXS J120328.8+024912 is the best candidate for a GRB X-ray afterglow 
simply due to the fact that the
ROSAT error box does not contain a bright ($m<22$ mag) stellar object (GRB host
galaxies are faint (e.g. Hogg $\&$ Fruchter 1999), though the light
curve is single-peaked. While it is difficult to determine the likelihood
that a flare of this large amplitude from a position with no optical
counterpart could be due to a statistical fluctuation, we note that this
event is among the largest amplitude events of our whole sample (see cols. 8
and 9 in Tab. \ref{xcand}). Also, the significance of the X-ray source itself
is huge (col. 4 in Tab. \ref{xcand}).

If we argue that the RASS data contain a few afterglows, then
data are obviously consistent with the expected theoretical rate
(especially considering the significant uncertainties
affecting our estimate of the afterglow expectation value). This
implies that GRB afterglows do not have a significantly wider 
beaming angle in the X-ray band relative to the gamma-ray band.
This is to some extent in agreement with predictions of
the ``standard'' fireball model
(Meszaros $\&$ Rees 1997; Piran 1999; Meszaros 1999), given the fact 
that we are only sampling
a few hours of emission following the GRB. As the fireball slows
due to interaction with a surrounding medium the bulk Lorentz 
factors of the flow decrease and the beaming angle increases.
However, the RASS data cover a time interval of $\sim$1--8 hrs
after the GRB event. During this time the fireball is expected to 
decelerate from $\Gamma \gax 100$ to $\Gamma \sim 10$. Thus, the flow
is still highly relativistic and the afterglow emission is still far 
from isotropic. One thus expects comparable detection rates for
prompt and delayed emission. 

On the other hand, if we argue that those of the events in Table 1 
which are not optically identified
are in fact GRB afterglows, then the rate apparently exceeds 
expectations. However, the enhancement factor is less than
a few. Furthermore, the uncertainties are large and the sample
is still small, thus the significance of this enhancement is small.
Again we would conclude that the RASS results support consistency
between observations and theoretical expectations, with only marginal
evidence for less beaming in the X-ray band. 

Both points of view basically conclude the same; beaming of GRBs and
of their afterglows is, if it exists, comparable. This conclusion
supports a similar result (Grindlay 1999) obtained from
an analysis of fast X-ray transients observed with {\it Ariel V}
(Pye $\&$ McHardy 1983) and earlier instruments. We also
emphasize that our results and those discussed by Grindlay 
(1999) can be used to place constraints on presently undetected 
GRB populations that preferentially emit in the X-ray band.
Dermer (1999) pointed out that the initial fireball Lorentz factor,
$\Gamma_0$, is crucial for determining the appearance of the GRB.
Since $\Gamma_0$ is related to the ratio of total burst energy to
rest mass energy of the baryon load a ``clean'' (low baryon load
and/or large energy) fireball is characterized by $\Gamma_0$ in
excess of 300 (according to Dermer's definition), while a ``dirty'' 
fireball (heavy load) is characterized by a very small Lorentz factor.
Dermer argues that clean fireballs produce GRBs of very short 
duration with emission predominantly in the high-energy regime, 
while dirty fireballs produce GRBs of long duration 
that preferentially radiate in the X-ray band. These bursts
are in fact predicted to be X-ray bright, but have probably not yet
been detected by BATSE and similar instruments, because these detectors 
are ``tuned'' to events for which $\Gamma_0$ falls in the range 200--400
(Dermer 1999). The absence of a significant number of X-ray 
transients in the RASS and the {\it Ariel} survey thus suggests that
the frequencies of ``dirty'' GRBs relative to bursts with a ``normal''
baryon load is comparable. 

Vietri {\it et al.} (1999) drew attention to the ``anomalous'' X-ray
afterglows from GRB 970508 and GRB 970828, which exhibit a resurgence of
soft X-ray emission and evidence for Fe-line emission. These authors 
interprete the delayed ``rebursts'' in the framework of the SupraNova
model (Vietri $\&$ Stella 1998) in which the GRB progenitor system 
creates a torus of iron-rich material. The GRB fireball heats the torus,
which cools via Bremsstrahlung, leading to a ``reburst'' in the X-ray
band. The emission pattern of this heated torus should be nearly isotropic,
so that one expects many X-ray afterglows that are not accompanied by 
GRBs. The RASS data place severe constraints on this type of reburst
scenario, because these delayed components are predicted (Vietri {\it et 
al.} 1999) to be bright (10$^{-4}$ erg cm$^{-2}$) and of long duration
($\sim$ 10$^3$ s). The rarity of afterglows in the RASS data suggests
that GRBs from ``SupraNovae'' do not constitute the bulk of the observed
GRB population, unless the GRBs are also roughly isotropic emitters (which
is in conflict with the correspondingly large energy requirements).  

Another constraint can be placed on GRBs related to supernovae (SN). If the 
association of GRB 980425 with SN1998bw is real (e.g. Galama \etal\ 1998,
Woosley \etal\ 1999) then such SN-related GRBs would dominate the total GRB 
rate by a factor of $\sim$1000 due to their low luminosities implied by the 
small redshift (z = 0.0085) of the host galaxy. 
It can be argued that GRB 980425 was beamed away from us, and we
merely saw the less beamed afterglow emission. If this is true,
we expect many X-ray afterglows in the RASS data. Again, our results
constrain these possibilities, but more quantitative results require 
detailed simulations that are beyond the scope of this paper.

In order to determine or further constrain differential beaming (X-rays
vs. $\gamma$-rays) on short timescales more sensitive surveys 
with larger exposures (in FOV and time) are needed. It would
also be important to establish the statistical properties of low energy
afterglows. In particular, for studies of this kind one needs better
knowledge of the distribution of peak X-ray fluxes and power law indices
of the temporal decays (to better estimate f in eq. 1).
BeppoSAX continues to provide these
measurements at an approximate rate of one afterglow every 1--2 month.
HETE2 will soon add events to this database, BATSE (in conjunction with
fast response systems on the ground such as LOTIS, ROTSE, and others) as well 
as AGILE, BALLERINA, GLAST, INTEGRAL, SWIFT, and perhaps other instruments
will provide this information
in the near to intermediate future. New insights and surprises
are likely to keep observers busy and theorists challenged. In the 
meantime the RASS observations presented here support 
the idea that early afterglow emission from GRBs has comparable
beaming properties in the X-ray and gamma-ray bands.

\begin{acknowledgements}
We are indebted to E. Costa and J. in 't Zand for providing X-ray afterglow 
light curves in digital form for Fig. 2.
JG and RS are supported by the German Bun\-des\-mi\-ni\-sterium f\"ur Bildung,
Wissenschaft, Forschung und Technologie
(BMBF/DLR) under contract Nos. 50 QQ 9602 3 and 50 OR 9708 6, respectively,
 and SZh by INTAS N 96-0315.
DHH expresses gratitude for support and hospitality during visits
to the AIP in Potsdam and the MPE in Garching.
The \ros\, project is supported by BMBF/DLR and the Max-Planck-Society.
This research has made use of the Simbad database, operated at CDS, 
Strasbourg, France and the Digitized Sky Survey (DSS) produced at 
the Space Telescope Science Institute under US Government grant NAG W-2166.

\end{acknowledgements}

\end{document}